\documentclass[aps,prl,showpacs,twocolumn,amsmath,amsmath,amssymb,amsfonts]{revtex4-1}
\usepackage{graphicx}
\usepackage{dcolumn}
\usepackage{bm}
\usepackage{color}
\usepackage{multirow}

\newcommand{\be}{\begin{equation}}
\newcommand{\ee}{\end{equation}}
\newcommand{\ba}{\begin{align}}
\newcommand{\ea}{\end{align}}
\newcommand{\sysb}{\left\{\begin{array}}
\newcommand{\syse}{\end{array}\right.}
\newcommand{\baa}{\begin{array}}
\newcommand{\eaa}{\end{array}}

\begin{document}
\title{Non-equilibrium absorbing state phase transitions in discrete-time quantum dynamics}
\author{Igor Lesanovsky}
\author{Katarzyna Macieszczak}
\author{Juan P. Garrahan}
\affiliation{School of Physics and Astronomy, University of Nottingham, Nottingham, NG7 2RD, UK}
\affiliation{Centre for the Mathematics and Theoretical Physics of Quantum Non-equilibrium Systems,
University of Nottingham, Nottingham NG7 2RD, UK}
\date{\today}
\begin{abstract}
We introduce a discrete-time quantum dynamics on a two-dimensional lattice that describes the evolution of a $1+1$-dimensional spin system. The underlying quantum map is constructed such that the reduced state at each time step is separable. We show that for long times this state becomes stationary and displays a continuous phase transition in the density of excited spins. This phenomenon can be understood through a connection to the so-called Domany-Kinzel automaton, which implements a classical non-equilibrium process that features a transition to an absorbing state. Near the transition density-density correlations become long-ranged, but interestingly the same is the case for quantum correlations despite the separability of the stationary state. We quantify quantum correlations through the local quantum uncertainty and show that in some cases they may be determined experimentally solely by measuring expectation values of classical observables. This work is inspired by recent experimental progress in the realization of Rydberg lattice quantum simulators, which --- in a rather natural way --- permit the realization of conditional quantum gates underlying the discrete-time dynamics discussed here.
\end{abstract}


\maketitle
\textit{Introduction ---} Recent years have witnessed breakthroughs in the realization of quantum simulator platforms based on cold atomic systems \cite{britton2012,Schauss_2015,labuhn2016,bernien2017,zhang2017}. One of the most recent generations of these quantum simulators is based on Rydberg atoms and offers freely programmable and addressable spin arrays \cite{Schauss_2015,labuhn2016,bernien2017,kim2017}. When excited to (high-lying) Rydberg states atoms interact strongly, thereby offering a very versatile platform for the study of quantum matter in an out of equilibrium.

Strong interactions between Rydberg atoms are moreover at the heart of implementations of quantum information processing protocols \cite{Ryd-QI} where they allow the realization of conditional gates \cite{Wilk2010,isenhower2010} that generate entangling operations. Digital quantum simulators \cite{buluta2009} employ such gates --- similar to the circuit-based approach to quantum computing --- and represent a route towards emulating quantum dynamics with exotic interactions. The possibility of digitally simulating open and closed many-body systems with Rydberg lattice systems was theoretically explored in Ref. \cite{weimer2010} and the capability of this platform for preparing exotic many-body systems and states was highlighted. While their experimental realization has not been achieved yet, first proof-of-principle demonstrations of the feasibility of this idea were demonstrated within a trapped ion quantum simulator \cite{lanyon2011} and superconducting circuits \cite{Salathe2015,barends2015}.

\begin{figure}[t!]
  \includegraphics[width=\columnwidth]{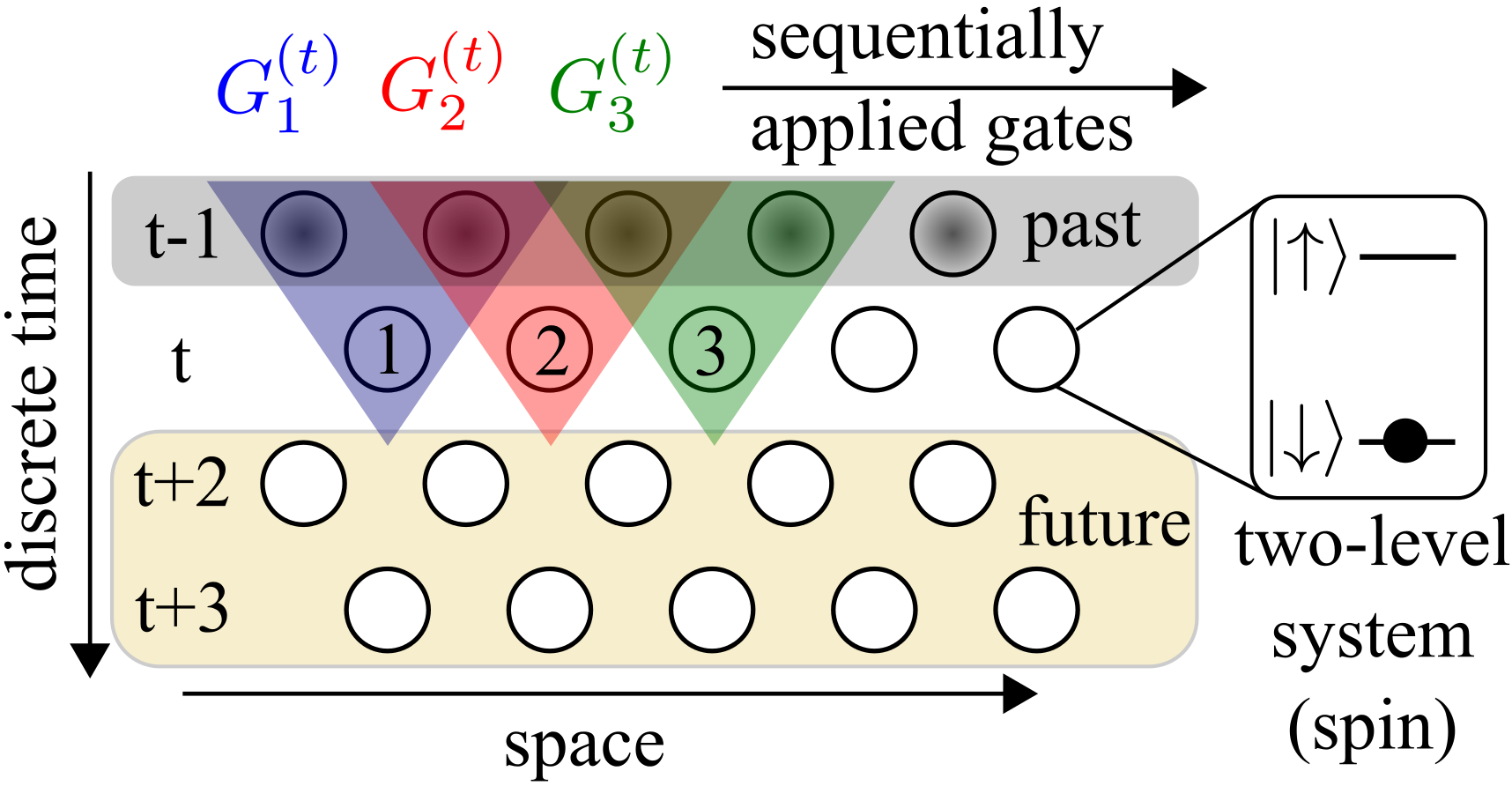}
  \caption{Two-dimensional ($1+1$) lattice system in which the horizontal (vertical) direction can be thought of as space (time). Each lattice site contains a single spin degree of freedom (for example encoded in an atom) which is initialized in the state $\left|\downarrow\right>$. An initial state is prepared on the first time slice and propagated towards future times, i.e. lower rows, by a sequence of gates that connect subsequent time slices. In the example here we use three-body gates $G_m^{(t)}$ which can be implemented for example in Rydberg lattice quantum simulators, where the spin degree of freedoms are encoded in two electronic levels.}
\label{fig:1}
\end{figure}
In this work we introduce a class of spin models with discrete-time quantum dynamics that lends itself rather naturally to the implementation on a Rydberg quantum simulator. The dynamics takes place within a $1+1$-dimensional lattice in which the directions can be thought of representing time and space, respectively. Propagation between time slices proceeds via the successive application of three-body gates that perform conditional rotations. Despite the fact that the dynamics of the whole system is unitary, the reduced state on the final time slice converges to a stationary state and may display a non-equilibrium phase transition. This stationary state features non-classical correlations that become long-ranged in the vicinity of the transition point. We illustrate our idea using an example that is efficiently solvable in the sense that it permits the mapping onto the non-equilibrium process of a classical cellular automaton for site percolation. Our work introduces a new aspect to ongoing attempts towards extending the concept of a cellular automaton into the quantum domain --- for a few examples see Refs. \cite{grossing1988,lent1993,meyer1996,gutschow2010,alonso2014,prosen2016,gopalakrishnan2018} --- and also connects to quantum generalizations of perceptrons in neural networks \cite{lewenstein1994,torrontegui2018}. Our proposed setting provides furthermore a natural testbed for assessing the capabilities of current Rydberg lattice quantum simulators: it possesses non-trivial features, such as a phase transition and long-ranged quantum correlations, but yet can be efficiently solved. It can thus be used for the certification of a Rydberg simulator in a regime (two dimensions, strong interactions, long times) which is usually numerically intractable.

\textit{The setting ---} The two-dimensional lattice system we are considering is depicted in Fig. \ref{fig:1}. Each row consists of $N$ sites, with a spin-$1/2$ degree of freedom per site. The horizontal and vertical directions we consider as space and time, respectively. The dynamics starts from a state where all spins are in the state $\left|\downarrow\right>$ except for the first time slice (first row) which is prepared in the desired initial configuration. The evolution then proceeds by applying a sequence of elementary gates linking the time slice at time $t$ to the time slice at time $t+1$.

For the case we are mainly interested in this work these elementary gates are unitary operators that act on three spins simultaneously --- two consecutive ones on time slice $t$ (control spins) and one on time slice $t+1$ (target spin), as shown in Fig. \ref{fig:1}. These gates perform a rotation of the state of the target spin, conditioned on the presence or absence of excited spins (in state $\left|\uparrow\right>$) among the two-control spins. We consider this type of gate here because it can be rather naturally implemented in Rydberg lattice quantum simulators as is discussed further below. Formally, we can write the gate as
\begin{eqnarray}
  G^{(t)}_m=P^{(t-1)}_{m, m+1}\otimes U^{(t)}_m + Q^{(t-1)}_{m, m+1}\otimes\mathbb{I}_m.\label{eq:gate}
\end{eqnarray}
Here $P^{(t-1)}_{m, m+1}$ and $Q^{(t-1)}_{m, m+1}$ are projection operators, which act on the control spins on time slice $t-1$ (with indices $m$ and $m+1$) and obey $P^{(t-1)}_{m, m+1}+Q^{(t-1)}_{m, m+1}=\mathbb{I}$. To be specific we use for now
\begin{eqnarray}
  P_{m, m+1}= 1-\left(1-n_m\right)\left(1-n_{m+1}\right),\label{eq:projector}
\end{eqnarray}
where $n_m=(1+\sigma^m_z)/2$, projects onto the excited state of the $m$-th spin on time slice $t-1$ and $\sigma^m_z$ is a Pauli matrix. The projector $P_{m, m+1}$ returns a non-zero value only if at least one of the control spins is in the excited state. When this is the case the unitary operator $U^{(t)}$ acts on the target atom on time slice $t$ and performs a spin rotation about the $y$-axis by an angle $\alpha$: $U=\exp\left(-i\frac{\alpha}{2}\sigma_y\right)$. Note, that we dropped the time slice index $t$ in the explicit forms of both the projectors and the unitary in order not to make the notation too contrived.

The rule (\ref{eq:gate}) can be considered as imposing a kinetic constraint in the dynamics, reminiscent of facilitated models of glasses \cite{Ritort2003}, such that local evolution only occurs if a certain condition is met. Constrained dynamics can give rise to complex evolution both in classical \cite{Garrahan2002} and in both closed and open quantum systems
\cite{Olmos2012,Horssen2015,Lan2017,Shiraishi2017,gopalakrishnan2018,Turner2017,Nandkishore2018}.
In particular, a rule akin to (\ref{eq:gate}), of at least one nearest neighbour in the excited state required to allow for local evolution, is known in classical facilitated models to lead to an effective dynamics of the reaction-diffusion kind \cite{Whitelam2004,Jack2006}, with the concomitant competition between active and inactive dynamical states.

In our model the propagation from time slice $t-1$ to $t$ is achieved via the concatenation of gates, $G^{(t)}_N...G^{(t)}_2 G^{(t)}_1$, where we assume periodic boundary conditions when applying $G^{(t)}_N$. Note, that due to the specific choice made in Eq. (\ref{eq:projector}) the actual order of the gates is not important since the projectors commute. The successive application of the gate $\mathcal{G}^{(t)}$ to subsequent time slices propagates the initial state and creates a pure state (provided that the initial state has been pure) on the entire lattice.

The reduced state $\rho_t$ on time slice $t$ is linked to the reduced state of the previous time slice by a recurrence relation:
\begin{eqnarray}
  \rho_{t}=\!\!\!\!\!\sum_{i_1,...i_N=1,2}\!\!\!\!\mathrm{Tr}\left[ X^{(i_1)}_1...\,X^{(i_N)}_N \rho_{t-1}\right] \rho^{(i_1)}\otimes...\otimes\rho^{(i_N)},\quad\label{eq:reduced_state}
\end{eqnarray}
with $X^{(1)}_m=P_{m, m+1}$ and $X^{(2)}_m=Q_{m, m+1}$ as well as $\rho^{(1)}=U\left|\downarrow\right>\left<\downarrow\right|U^\dagger$ and
$\rho^{(2)}=\left|\downarrow\right>\left<\downarrow\right|$. The state $\rho_t$ is thus separable and formed by a convex superposition of product states of the form  $\rho^{(i_1)}\otimes...\otimes\rho^{(i_N)}$. The weight of each state is given by the expectation value of the product of projection operators taken in the state of the previous time slice, $\rho_{t-1}$. In our protocol local quantum operations, such as $U^{(t)}_m$, are conditioned by a measurement result $i_m$, that can be communicated "classically". Such scheme cannot produce an entangled state on time slice $t$. Nevertheless, $\rho_t$ can exhibit non-classical correlations as we show later.

\textit{Mean field approximation ---} In order to gain a first understanding of the discrete-time dynamics we conduct a mean field study. To this end we consider the evolution of the local density on site $m$ under the gate (\ref{eq:gate}), which yields
\begin{eqnarray*}
  \langle n_m^{(t)} \rangle_t=\langle\downarrow\mid {G^{(t)}_m}^\dagger n^{(t)}_m G^{(t)}_m \mid\downarrow\rangle_t=x\,P^{(t-1)}_{m,m+1},
\end{eqnarray*}
with $x=\langle\downarrow\mid {U^{(t)}_m}^\dagger n^{(t)}_m U^{(t)}_m \mid\downarrow\rangle_t=\sin^2\left(\frac{\alpha}{2}\right)\,\epsilon \, [0,1]$. We take the expectation value over the $t-1$-time slice, make use of the form (\ref{eq:projector}) of the projector $P_{m,m+1}$ and perform the mean field approximation (decoupling of pair correlators and assumption of homogeneity). This yields a recurrence relation, connecting the mean field densities $\nu$ at time slices $t$ and $t-1$: $\nu^{(t)}=x\,\nu^{(t-1)}\left(2-\nu^{(t-1)}\right)$. To make progress we turn the recurrence relation into a differential equation [$\nu^{(t)}\rightarrow \nu(t)$, $\nu^{(t)}-\nu^{(t-1)}\rightarrow \partial_t \nu(t)$]. Choosing the initial condition $n(0)=1$, we obtain the solution
$\nu(t)=(2x-1)/\left[x+\exp\left(t\left[1-2x\right]\right)\left(x-1\right)\right]$,
which has an interesting limiting behavior at long times: for $x<x_\mathrm{crit}=1/2$ we find $\lim_{t\rightarrow\infty}\nu(t)=0$, while for $x>x_\mathrm{crit}$ the excitation density assumes the non-zero stationary value $\lim_{t\rightarrow\infty}\nu(t)=(x-x_\mathrm{crit})/(2x)$. Thus, $x_\mathrm{crit}$ defines a critical rotation angle $\alpha_\mathrm{crit}=\pi/4$ which in the limit $t\rightarrow\infty$ separates two qualitatively different states. At $x=x_\mathrm{crit}$ we find $\nu(t)=2/(2+t)$ and thus the density displays an algebraic approach to stationarity. This result is reminiscent of mean field calculations of classical reaction-diffusion problems that feature absorbing state phase transitions \cite{hinrichsen2000}.

\textit{Mapping to a classical non-equilibrium process ---} Further insight into this phase transition behavior can be obtained by realizing that there is a link to a classical stochastic process. The reason is that, due to the separability of the reduced density matrices $\rho_t$ and the structure of the projectors (\ref{eq:projector}), the probabilities $\mathrm{Tr}\left[ X^{(i_1)}_1...X^{(i_N)}_N \rho_{t-1}\right]$, which appear in the reduced state (\ref{eq:reduced_state}), can be generated via a classical discrete time dynamics. Like the quantum dynamics this process takes place on a two-dimensional lattice, as depicted in Fig. \ref{fig:1}, that contains classical spins (which are either up or down), initially prepared in the state $\|\!\downarrow\rangle\!\rangle$. The discrete time evolution proceeds via the classical maps
\begin{eqnarray}
 W^{(t)}_m= P^{(t-1)}_{m, m+1}\otimes \left(
                                        \begin{array}{cc}
                                         1-x & x \\
                                          x & 1-x \\
                                        \end{array}
                                      \right)_m
  \!\!\!\!+ Q^{(t-1)}_{m, m+1}\otimes\mathbb{I}_m\quad\label{eq:DK_map}
\end{eqnarray}
which are applied on a probability vector in order to propagate the system between time slices. This dynamics implements an instance of the so-called Domany-Kinzel (DK) cellular automaton \cite{domany1984,hinrichsen2000} and it performs a flip of the target spin (time slice $t$) with probability $x$, provided that the projection operator $P^{(t-1)}_{m, m+1}$ yields a non-zero value when applied to the control spin on time slice $t-1$. Under this dynamics the reduced probability vector $\|\mathbf{p}\rangle\!\rangle_t$ of time slice $t$ evolves according to
\begin{align}
  \|\mathbf{p}\rangle\!\rangle_t=\!\!\!\!\!\!\sum_{i_1,...i_N=1,2} \!\!\!\!\langle\!\langle+\|X^{(i_1)}_1...\,X^{(i_N)}_N \|\mathbf{p}\rangle\!\rangle_{t-1}\|s_{i_1}\rangle\!\rangle\otimes...\otimes \|s_{i_N}\rangle\!\rangle\label{eq:reduced_probability}
\end{align}
with $\|s_{1}\rangle\!\rangle=\|\!\uparrow\rangle\!\rangle$ and $\|s_{2}\rangle\!\rangle=(1-x)\|\!\downarrow\rangle\!\rangle+x\|\!\uparrow\rangle\!\rangle$. Note, that instead of taking the trace, expectation values in this classical description are calculated by applying the desired operator to the probability vector and multiplying from the left with a (flat) reference state: for $N$ spins this is $\|+\rangle\!\rangle=\bigotimes_{m=1}^N \left[\|\!\downarrow\rangle\!\rangle_m+\|\!\uparrow\rangle\!\rangle_m\right]$.

The structural resemblance between the reduced state (\ref{eq:reduced_state}) and the probability vector (\ref{eq:reduced_probability}) is evident. The local quantum states $\rho^{(k)}$ and classical states $\|s_{k}\rangle\!\rangle$ are constructed such that they yield the same expectation values for classical observables, e.g.
$\mathrm{Tr}\left(n\, \rho^{(m)}\right) = \langle\!\langle+\|n\|s_{m}\rangle\!\rangle = x\,\delta_{m,1}$. Thus, also the states (\ref{eq:reduced_state}) and (\ref{eq:reduced_probability}) yield identical expectation values of classical observables, and in this sense the discrete time quantum dynamics is mapped onto a classical process.

\begin{figure}[t!]
  \includegraphics[width=0.8\columnwidth]{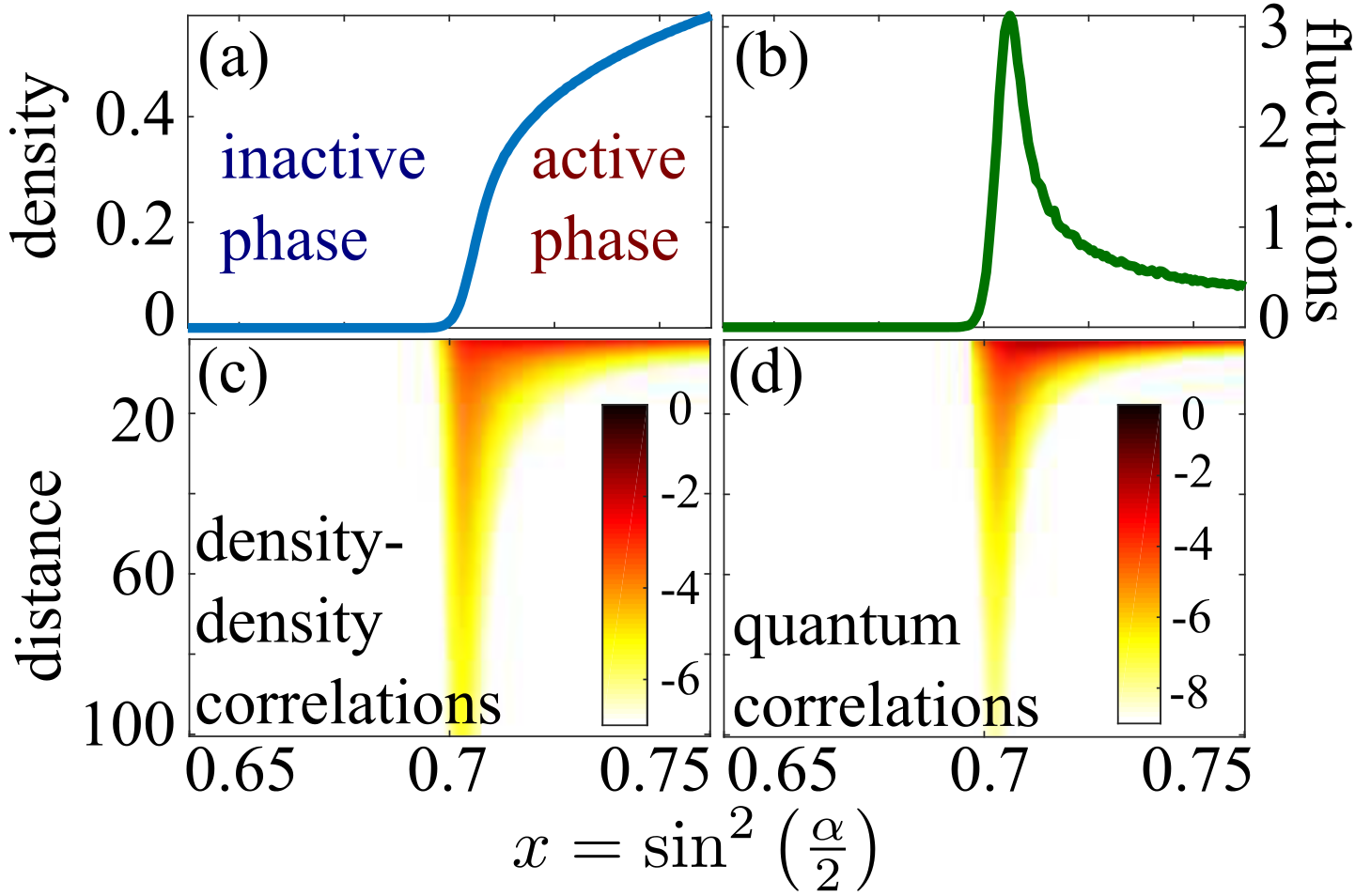}
  \caption{Density, fluctuations and quantum correlations (numerical simulations for $N=1000$, $7000$ time steps and $5000$ averages). (a) The mean density $\left<n\right>=\frac{1}{N}\sum_m \left<n_m\right>$ displays a phase transition at $x_\mathrm{crit}\approx 0.7$, from an inactive (zero density) to an active phase (finite density). This transition belongs to the directed percolation universality class. (b) At the critical point the fluctuations, $\frac{(\Delta N)^2}{N}=\frac{1}{N}\left[\sum_{ij}\left<n_i n_j\right>- N^2 \left<n\right> \right]$, exhibit a pronounced peak. (c) At the phase transition the (connected) density-density correlations $C_{ij}=c^2_{ij}-\left<n\right>^2$ become long-ranged. The density plot shows the natural logarithm of $C_{ij}$ as a function of the distance $|i-j|$. (d) Quantum correlations, quantified through the local quantum uncertainty (LQU), also become long-range ranged in the vicinity of the critical point. The density plot shows the natural logarithm of the LQU of the reduced two-spin density matrix $\rho_{ij}$, as a function of the distance $|i-j|$.}
\label{fig:2}
\end{figure}

The connection to the DK cellular automaton provides an explanation for the phase transition behavior observed in the mean field calculation: it is known that the cellular automaton dynamics (\ref{eq:DK_map}) leads to a non-equilibrium stationary state which displays a continuous (absorbing state) phase transition between a so-called inactive phase --- in which the expectation value of the average density $\langle n \rangle=\frac{1}{N} \sum_m^N \langle n_m \rangle$ is zero --- and an active phase in which $\langle n \rangle\neq 1$. This transition occurs at $x\approx 0.7$ and is in the directed percolation universality class. The corresponding numerical data is shown in Fig. \ref{fig:2}(a-c).

\textit{Quantum correlations ---} Despite being separable and related to a classical dynamics, the state (\ref{eq:reduced_state}) possesses non-classical correlations, as we show now. Furthermore, by exploiting the mapping to the DK cellular automaton dynamics we find that it is possible to extract quantum correlations from the measurement of classical observables, which are straight-forwardly accessible on Rydberg quantum simulators \cite{labuhn2016}.

As a measure for quantum correlations we employ the local quantum uncertainty (LQU) put forward in Ref. \cite{girolami2013} which is a variant of bipartite quantum discord \cite{Zurek2000,Ollivier2001,Henderson2001}. It quantifies how much of the statistical error of a local measurement is due to the non-commutativity between the state and the measured observable, which is caused by the state's coherence. By minimising over the choice of the observable, only non-local coherences, corresponding to quantum correlations, are captured. For the reduced state $\rho_{ij}$ of two spins the LQU is defined as $\ell_{ij}=1-\lambda_\mathrm{max}\left\{W^{ij}\right\}$, where $\lambda_\mathrm{max}\left\{W^{ij}\right\}$ is the largest eigenvalue of the matrix $W^{ij}$ with components $  W^{ij}_{\alpha\beta}=\mathrm{Tr}\left(\rho^{1/2}_{ij}\sigma^i_\alpha\rho^{1/2}_{ij}\sigma^i_\beta\right)$. The reduced density matrix $\rho_{ij}$ can be obtained entirely from measuring the local density and density-density correlations between sites $i$ and $j$. To see this we exploit the special structure of the reduced state (\ref{eq:reduced_state}): each term of the sum contains a product of pure states which allows to relate expectation values of off-diagonal operators to those of diagonal observables, e.g. $\mathrm{Tr} \left(\sigma^\pm_i\rho_t\right)= \sqrt{\frac{1-x}{x}}\mathrm{Tr}\left(n_i\rho_t\right)=\sqrt{\frac{1-x}{x}}\left<n_i\right>$. Using this property, and assuming translation invariance ($\left<n\right>=\left<n_i\right>=\left<n_j\right>$), one obtains
\begin{eqnarray}
  \rho_{ij}&=&\left(
              \begin{array}{cc}
                c_{ij} & x\, c_{ij} \\
                x\, c_{ij} & 1-c_{ij} \\
              \end{array}
            \right)\otimes \left(
              \begin{array}{cc}
                c_{ij} & x\, c_{ij} \\
                x\, c_{ij} & 1-c_{ij} \\
              \end{array}
            \right)\nonumber\\
            &&+\left[\left<n\right>-c_{ij}\right]\left(
                                       \begin{array}{cccc}
                                         0 & 0 & 0 & 0 \\
                                         0 & 1 & 0 & x \\
                                         0 & 0 & 1 & x \\
                                         0 & x & x & -2 \\
                                       \end{array}
                                     \right).\label{eq:two-spin_state}
\end{eqnarray}
with $c_{ij}=\sqrt{\left<n_in_j\right>}$ being the square root of the density-density correlation function.

In the absence of correlations one has $c_{ij}=\left<n\right>$. Here, the second term in Eq. (\ref{eq:two-spin_state}) vanishes and $\rho_{ij}$ becomes a product state without quantum correlations. This is the case away from a phase transition where correlations between two sites are decaying rapidly as a function of their distance. Near a phase transition, however, correlations are long-ranged, as is shown in Fig. \ref{fig:2}(c), where we display the connected density-density correlation function $C_{ij}=c^2_{ij}-\left<n\right>^2$. Here also finite and long-ranged quantum correlations, characterized through the LQU, emerge, as can be seen in Fig. \ref{fig:2}(d).

\begin{figure}[t!]
  \includegraphics[width=\columnwidth]{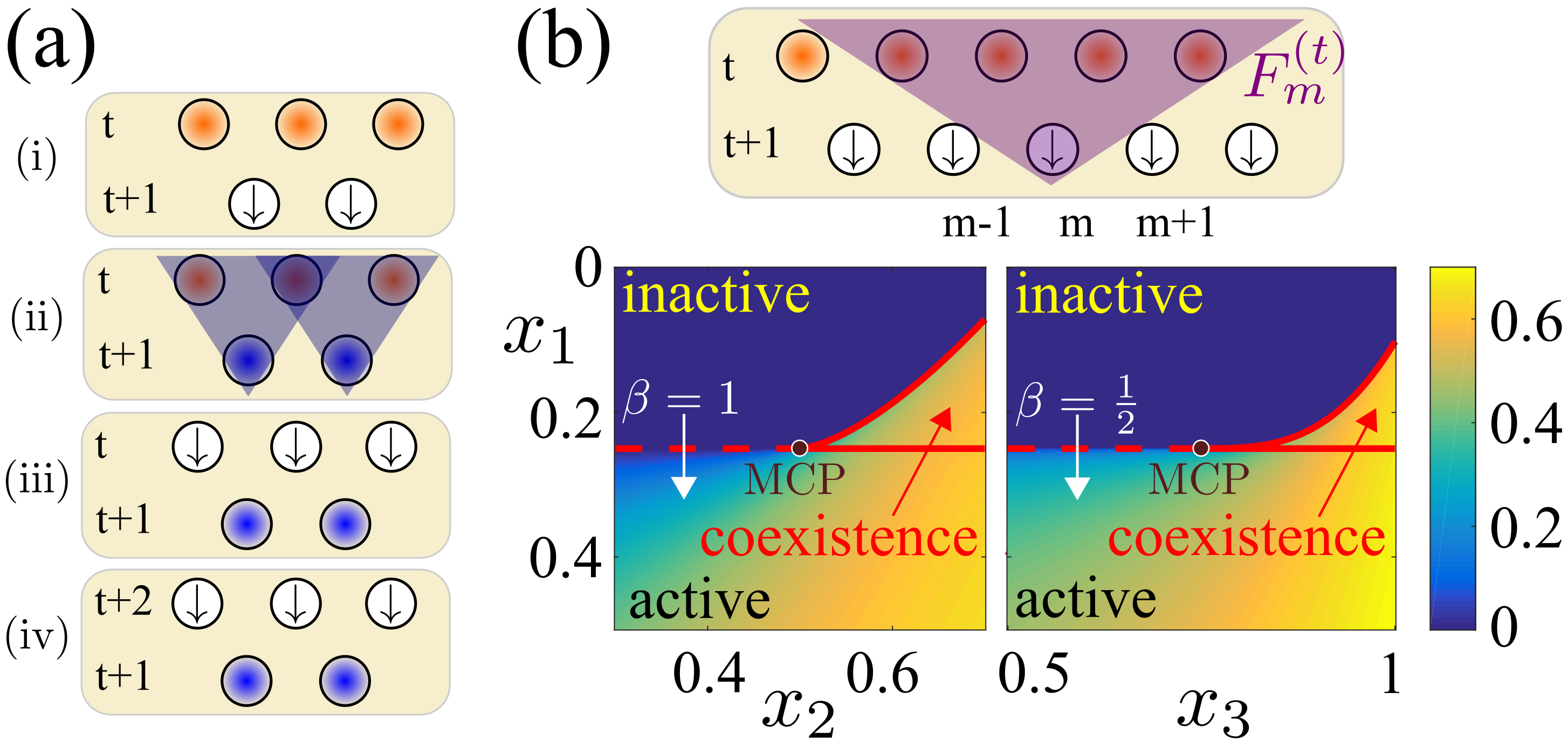}
  \caption{(a) Implementation of $1+1$-discrete-time dynamics with two spin chains. For further details see main text. (b) A generalization of the underlying gate operation to four source atoms [Eq. (\ref{eq:gen_gate}) with $K=4$] allows to implement non-equilibrium processes which features a variety of absorbing state phase transitions. Shown are cuts through the mean field phase diagram of Eq. (\ref{eq:gen_mf}). The dashed lines correspond to continuous phase transitions which terminate in the multi-critical point (MCP) at $\{x^\mathrm{crit}_1,x^\mathrm{crit}_2,x^\mathrm{crit}_3\}=\{1/4,1/2,3/4\}$.
  Upon crossing the dashed lines the mean field density shows scaling behavior of the form $\nu \sim (x_\alpha-x^\mathrm{crit}_\alpha)^\beta$ with $\beta$ being the static critical exponent. Solid lines demarcate regions in which an active and inactive phase coexist.}
\label{fig:3}
\end{figure}
\textit{Implementation with Rydberg atoms ---} The open cellular automaton model discussed here can be implemented on Rydberg quantum simulators \cite{labuhn2016,bernien2017}. The three-body gates underlying the gate (\ref{eq:gate}) are implemented by employing the blockade interaction \cite{Wilk2010} which yields conditional unitaries \cite{ostmann2017} discriminating between the cases in which at least one or none of the source atom is excited. For the experimental investigation of the non-equilibrium dynamics it is moreover not necessary to have a two-dimensional lattice. Two parallel one-dimensional arrays are sufficient for the following protocol [see also Fig. \ref{fig:3}(a)]: (i) The initial state is prepared on the first chain and all sites of the second chain are prepared in the state $\left|\downarrow\right>$. (ii) The discrete-time propagation is performed from the first to the second chain. (iii) The first chain is reset, so that all sites are in state $\left|\downarrow\right>$. (iv) The process is repeated but the role of the chains is interchanged.

\textit{Generalizations and future perspective ---} Generalizations of the dynamics presented here can be achieved by extending the fundamental gate (\ref{eq:gate}) to more source/target atoms and/or by introducing more conditional spin rotations. One possible extension of the gate to $K$ source atoms and one target atom is given by
\begin{eqnarray}
  F^{(t)}_m=\sum_{k=0}^K \Pi^{(t-1)}_m(k,K) \otimes U^{(t)}_m(\alpha_k).\label{eq:gen_gate}
\end{eqnarray}
Here the operators $\Pi_m(k,K)$ project on the subspace containing $k$ excitations among the $K$ source atoms whose state conditions the state change of the $m$-th target atom. The latter is rotated by the unitary $U(\alpha_k)=\exp\left(-i\frac{\alpha_k}{2}\sigma_y\right)$. We anticipate two interesting cases here:

(i) $K=2$ source sites and rotation angles are given by $\alpha_2=\pi$, $\alpha_1=\alpha$ and $\alpha_0=0$: The corresponding non-equilibrium process has the two absorbing states $\left|\downarrow\downarrow...\downarrow\right>$ and $\left|\uparrow\uparrow...\uparrow\right>$. At $\alpha=\pi/2$ the stationary state switches between these two possibilities and displays a phase transition that is in the directed compact percolation universality class \cite{essam1989}.

(ii) $\alpha_0=0$, which ensures the presence of the absorbing state $\left|\downarrow\downarrow...\downarrow\right>$: Performing a mean field treatment based on the gate (\ref{eq:gen_gate}) we find that the mean field density follows the recurrence relation
\begin{eqnarray}
  \nu^{(t)}=\sum_{k=1}^K x_k\binom{K}{k} \left(\nu^{(t-1)}\right)^k\left(1-\nu^{(t-1)}\right)^{K-k},\label{eq:gen_mf}
\end{eqnarray}
where $x_k=\sin^2\left(\frac{\alpha_k}{2}\right)$. This process features a host of absorbing state phase transitions, coexistence regions and critical lines. Moreover, a suitable choice of the rotation angles $\alpha_k$ allows to set all terms of order smaller than $K$ to zero which tunes the system to a multi-critical point (similar to tri-critical directed percolation \cite{lubeck2006}): $\nu^{(t)}-\nu^{(t-1)}\propto - \left(\nu^{(t-1)}\right)^K$. Here the mean field density displays a power-law behavior on approach to stationarity: $\nu(t)\sim t^{1/(1-K)}$. In Fig. \ref{fig:3}(b) we illustrate the case $K=4$.

A interesting subject for future investigations is the realization of non-equilibrium processes with absorbing (dark-)states \cite{roscher2018} that feature entanglement and/or phase coherence between different sites. Those can be achieved by employing projectors in the fundamental gate (\ref{eq:gen_gate}) that project for example on two-site entangled states, in conjunction with unitary operations acting on two and more target sites.

\emph{Acknowledgments ---} We thank M. Marcuzzi and M. M\"uller for useful discussions. The research  leading  to  these  results  has  received  funding  from the European Research Council under the European Union’s Seventh Framework Programme (FP/2007-2013)/ERC Grant Agreement No.
335266 (ESCQUMA), the EPSRC Grant No. EP/M014266/1, and the H2020-FETPROACT-2014 Grant No. 640378 (RYSQ). I.L. gratefully acknowledges funding through the Royal Society Wolfson Research Merit Award.


\begin{thebibliography}{43}%
\makeatletter
\providecommand \@ifxundefined [1]{%
 \@ifx{#1\undefined}
}%
\providecommand \@ifnum [1]{%
 \ifnum #1\expandafter \@firstoftwo
 \else \expandafter \@secondoftwo
 \fi
}%
\providecommand \@ifx [1]{%
 \ifx #1\expandafter \@firstoftwo
 \else \expandafter \@secondoftwo
 \fi
}%
\providecommand \natexlab [1]{#1}%
\providecommand \enquote  [1]{``#1''}%
\providecommand \bibnamefont  [1]{#1}%
\providecommand \bibfnamefont [1]{#1}%
\providecommand \citenamefont [1]{#1}%
\providecommand \href@noop [0]{\@secondoftwo}%
\providecommand \href [0]{\begingroup \@sanitize@url \@href}%
\providecommand \@href[1]{\@@startlink{#1}\@@href}%
\providecommand \@@href[1]{\endgroup#1\@@endlink}%
\providecommand \@sanitize@url [0]{\catcode `\\12\catcode `\$12\catcode
  `\&12\catcode `\#12\catcode `\^12\catcode `\_12\catcode `\%12\relax}%
\providecommand \@@startlink[1]{}%
\providecommand \@@endlink[0]{}%
\providecommand \url  [0]{\begingroup\@sanitize@url \@url }%
\providecommand \@url [1]{\endgroup\@href {#1}{\urlprefix }}%
\providecommand \urlprefix  [0]{URL }%
\providecommand \Eprint [0]{\href }%
\providecommand \doibase [0]{http://dx.doi.org/}%
\providecommand \selectlanguage [0]{\@gobble}%
\providecommand \bibinfo  [0]{\@secondoftwo}%
\providecommand \bibfield  [0]{\@secondoftwo}%
\providecommand \translation [1]{[#1]}%
\providecommand \BibitemOpen [0]{}%
\providecommand \bibitemStop [0]{}%
\providecommand \bibitemNoStop [0]{.\EOS\space}%
\providecommand \EOS [0]{\spacefactor3000\relax}%
\providecommand \BibitemShut  [1]{\csname bibitem#1\endcsname}%
\let\auto@bib@innerbib\@empty
\bibitem [{\citenamefont {Britton}\ \emph {et~al.}(2012)\citenamefont
  {Britton}, \citenamefont {Sawyer}, \citenamefont {Keith}, \citenamefont
  {Wang}, \citenamefont {Freericks}, \citenamefont {Uys}, \citenamefont
  {Biercuk},\ and\ \citenamefont {Bollinger}}]{britton2012}%
  \BibitemOpen
  \bibfield  {author} {\bibinfo {author} {\bibfnamefont {J.~W.}\ \bibnamefont
  {Britton}}, \bibinfo {author} {\bibfnamefont {B.~C.}\ \bibnamefont {Sawyer}},
  \bibinfo {author} {\bibfnamefont {A.~C.}\ \bibnamefont {Keith}}, \bibinfo
  {author} {\bibfnamefont {C.-C.~J.}\ \bibnamefont {Wang}}, \bibinfo {author}
  {\bibfnamefont {J.~K.}\ \bibnamefont {Freericks}}, \bibinfo {author}
  {\bibfnamefont {H.}~\bibnamefont {Uys}}, \bibinfo {author} {\bibfnamefont
  {M.~J.}\ \bibnamefont {Biercuk}}, \ and\ \bibinfo {author} {\bibfnamefont
  {J.~J.}\ \bibnamefont {Bollinger}},\ }\href@noop {} {\bibfield  {journal}
  {\bibinfo  {journal} {Nature}\ }\textbf {\bibinfo {volume} {484}},\ \bibinfo
  {pages} {489} (\bibinfo {year} {2012})}\BibitemShut {NoStop}%
\bibitem [{\citenamefont {Schau{\ss}}\ \emph {et~al.}(2015)\citenamefont
  {Schau{\ss}}, \citenamefont {Zeiher}, \citenamefont {Fukuhara}, \citenamefont
  {Hild}, \citenamefont {Cheneau}, \citenamefont {Macr{\`\i}}, \citenamefont
  {Pohl}, \citenamefont {Bloch},\ and\ \citenamefont
  {Gro{\ss}}}]{Schauss_2015}%
  \BibitemOpen
  \bibfield  {author} {\bibinfo {author} {\bibfnamefont {P.}~\bibnamefont
  {Schau{\ss}}}, \bibinfo {author} {\bibfnamefont {J.}~\bibnamefont {Zeiher}},
  \bibinfo {author} {\bibfnamefont {T.}~\bibnamefont {Fukuhara}}, \bibinfo
  {author} {\bibfnamefont {S.}~\bibnamefont {Hild}}, \bibinfo {author}
  {\bibfnamefont {M.}~\bibnamefont {Cheneau}}, \bibinfo {author} {\bibfnamefont
  {T.}~\bibnamefont {Macr{\`\i}}}, \bibinfo {author} {\bibfnamefont
  {T.}~\bibnamefont {Pohl}}, \bibinfo {author} {\bibfnamefont {I.}~\bibnamefont
  {Bloch}}, \ and\ \bibinfo {author} {\bibfnamefont {C.}~\bibnamefont
  {Gro{\ss}}},\ }\href@noop {} {\bibfield  {journal} {\bibinfo  {journal}
  {Science}\ }\textbf {\bibinfo {volume} {347}},\ \bibinfo {pages} {1455}
  (\bibinfo {year} {2015})}\BibitemShut {NoStop}%
\bibitem [{\citenamefont {Labuhn}\ \emph {et~al.}(2016)\citenamefont {Labuhn},
  \citenamefont {Barredo}, \citenamefont {Ravets}, \citenamefont
  {De~L{\'e}s{\'e}leuc}, \citenamefont {Macr{\`\i}}, \citenamefont {Lahaye},\
  and\ \citenamefont {Browaeys}}]{labuhn2016}%
  \BibitemOpen
  \bibfield  {author} {\bibinfo {author} {\bibfnamefont {H.}~\bibnamefont
  {Labuhn}}, \bibinfo {author} {\bibfnamefont {D.}~\bibnamefont {Barredo}},
  \bibinfo {author} {\bibfnamefont {S.}~\bibnamefont {Ravets}}, \bibinfo
  {author} {\bibfnamefont {S.}~\bibnamefont {De~L{\'e}s{\'e}leuc}}, \bibinfo
  {author} {\bibfnamefont {T.}~\bibnamefont {Macr{\`\i}}}, \bibinfo {author}
  {\bibfnamefont {T.}~\bibnamefont {Lahaye}}, \ and\ \bibinfo {author}
  {\bibfnamefont {A.}~\bibnamefont {Browaeys}},\ }\href@noop {} {\bibfield
  {journal} {\bibinfo  {journal} {Nature}\ }\textbf {\bibinfo {volume} {534}},\
  \bibinfo {pages} {667} (\bibinfo {year} {2016})}\BibitemShut {NoStop}%
\bibitem [{\citenamefont {Bernien}\ \emph {et~al.}(2017)\citenamefont
  {Bernien}, \citenamefont {Schwartz}, \citenamefont {Keesling}, \citenamefont
  {Levine}, \citenamefont {Omran}, \citenamefont {Pichler}, \citenamefont
  {Choi}, \citenamefont {Zibrov}, \citenamefont {Endres}, \citenamefont
  {Greiner} \emph {et~al.}}]{bernien2017}%
  \BibitemOpen
  \bibfield  {author} {\bibinfo {author} {\bibfnamefont {H.}~\bibnamefont
  {Bernien}}, \bibinfo {author} {\bibfnamefont {S.}~\bibnamefont {Schwartz}},
  \bibinfo {author} {\bibfnamefont {A.}~\bibnamefont {Keesling}}, \bibinfo
  {author} {\bibfnamefont {H.}~\bibnamefont {Levine}}, \bibinfo {author}
  {\bibfnamefont {A.}~\bibnamefont {Omran}}, \bibinfo {author} {\bibfnamefont
  {H.}~\bibnamefont {Pichler}}, \bibinfo {author} {\bibfnamefont
  {S.}~\bibnamefont {Choi}}, \bibinfo {author} {\bibfnamefont {A.~S.}\
  \bibnamefont {Zibrov}}, \bibinfo {author} {\bibfnamefont {M.}~\bibnamefont
  {Endres}}, \bibinfo {author} {\bibfnamefont {M.}~\bibnamefont {Greiner}},
  \emph {et~al.},\ }\href@noop {} {\bibfield  {journal} {\bibinfo  {journal}
  {Nature}\ }\textbf {\bibinfo {volume} {551}},\ \bibinfo {pages} {579}
  (\bibinfo {year} {2017})}\BibitemShut {NoStop}%
\bibitem [{\citenamefont {Zhang}\ \emph {et~al.}(2017)\citenamefont {Zhang},
  \citenamefont {Pagano}, \citenamefont {Hess}, \citenamefont {Kyprianidis},
  \citenamefont {Becker}, \citenamefont {Kaplan}, \citenamefont {Gorshkov},
  \citenamefont {Gong},\ and\ \citenamefont {Monroe}}]{zhang2017}%
  \BibitemOpen
  \bibfield  {author} {\bibinfo {author} {\bibfnamefont {J.}~\bibnamefont
  {Zhang}}, \bibinfo {author} {\bibfnamefont {G.}~\bibnamefont {Pagano}},
  \bibinfo {author} {\bibfnamefont {P.~W.}\ \bibnamefont {Hess}}, \bibinfo
  {author} {\bibfnamefont {A.}~\bibnamefont {Kyprianidis}}, \bibinfo {author}
  {\bibfnamefont {P.}~\bibnamefont {Becker}}, \bibinfo {author} {\bibfnamefont
  {H.}~\bibnamefont {Kaplan}}, \bibinfo {author} {\bibfnamefont {A.~V.}\
  \bibnamefont {Gorshkov}}, \bibinfo {author} {\bibfnamefont {Z.-X.}\
  \bibnamefont {Gong}}, \ and\ \bibinfo {author} {\bibfnamefont
  {C.}~\bibnamefont {Monroe}},\ }\href@noop {} {\bibfield  {journal} {\bibinfo
  {journal} {Nature}\ }\textbf {\bibinfo {volume} {551}},\ \bibinfo {pages}
  {601} (\bibinfo {year} {2017})}\BibitemShut {NoStop}%
\bibitem [{\citenamefont {Kim}\ \emph {et~al.}(2017)\citenamefont {Kim},
  \citenamefont {Park}, \citenamefont {Kim}, \citenamefont {Sim},\ and\
  \citenamefont {Ahn}}]{kim2017}%
  \BibitemOpen
  \bibfield  {author} {\bibinfo {author} {\bibfnamefont {H.}~\bibnamefont
  {Kim}}, \bibinfo {author} {\bibfnamefont {Y.}~\bibnamefont {Park}}, \bibinfo
  {author} {\bibfnamefont {K.}~\bibnamefont {Kim}}, \bibinfo {author}
  {\bibfnamefont {H.-S.}\ \bibnamefont {Sim}}, \ and\ \bibinfo {author}
  {\bibfnamefont {J.}~\bibnamefont {Ahn}},\ }\href@noop {} {\bibfield
  {journal} {\bibinfo  {journal} {arXiv:1712.02065}\ } (\bibinfo {year}
  {2017})}\BibitemShut {NoStop}%
\bibitem [{\citenamefont {Saffman}\ \emph {et~al.}(2010)\citenamefont
  {Saffman}, \citenamefont {Walker},\ and\ \citenamefont {M\o{}lmer}}]{Ryd-QI}%
  \BibitemOpen
  \bibfield  {author} {\bibinfo {author} {\bibfnamefont {M.}~\bibnamefont
  {Saffman}}, \bibinfo {author} {\bibfnamefont {T.~G.}\ \bibnamefont {Walker}},
  \ and\ \bibinfo {author} {\bibfnamefont {K.}~\bibnamefont {M\o{}lmer}},\
  }\href@noop {} {\bibfield  {journal} {\bibinfo  {journal} {Rev. Mod. Phys.}\
  }\textbf {\bibinfo {volume} {82}},\ \bibinfo {pages} {2313} (\bibinfo {year}
  {2010})}\BibitemShut {NoStop}%
\bibitem [{\citenamefont {Wilk}\ \emph {et~al.}(2010)\citenamefont {Wilk},
  \citenamefont {Ga\"etan}, \citenamefont {Evellin}, \citenamefont {Wolters},
  \citenamefont {Miroshnychenko}, \citenamefont {Grangier},\ and\ \citenamefont
  {Browaeys}}]{Wilk2010}%
  \BibitemOpen
  \bibfield  {author} {\bibinfo {author} {\bibfnamefont {T.}~\bibnamefont
  {Wilk}}, \bibinfo {author} {\bibfnamefont {A.}~\bibnamefont {Ga\"etan}},
  \bibinfo {author} {\bibfnamefont {C.}~\bibnamefont {Evellin}}, \bibinfo
  {author} {\bibfnamefont {J.}~\bibnamefont {Wolters}}, \bibinfo {author}
  {\bibfnamefont {Y.}~\bibnamefont {Miroshnychenko}}, \bibinfo {author}
  {\bibfnamefont {P.}~\bibnamefont {Grangier}}, \ and\ \bibinfo {author}
  {\bibfnamefont {A.}~\bibnamefont {Browaeys}},\ }\href {\doibase
  10.1103/PhysRevLett.104.010502} {\bibfield  {journal} {\bibinfo  {journal}
  {Phys. Rev. Lett.}\ }\textbf {\bibinfo {volume} {104}},\ \bibinfo {pages}
  {010502} (\bibinfo {year} {2010})}\BibitemShut {NoStop}%
\bibitem [{\citenamefont {Isenhower}\ \emph {et~al.}(2010)\citenamefont
  {Isenhower}, \citenamefont {Urban}, \citenamefont {Zhang}, \citenamefont
  {Gill}, \citenamefont {Henage}, \citenamefont {Johnson}, \citenamefont
  {Walker},\ and\ \citenamefont {Saffman}}]{isenhower2010}%
  \BibitemOpen
  \bibfield  {author} {\bibinfo {author} {\bibfnamefont {L.}~\bibnamefont
  {Isenhower}}, \bibinfo {author} {\bibfnamefont {E.}~\bibnamefont {Urban}},
  \bibinfo {author} {\bibfnamefont {X.}~\bibnamefont {Zhang}}, \bibinfo
  {author} {\bibfnamefont {A.}~\bibnamefont {Gill}}, \bibinfo {author}
  {\bibfnamefont {T.}~\bibnamefont {Henage}}, \bibinfo {author} {\bibfnamefont
  {T.~A.}\ \bibnamefont {Johnson}}, \bibinfo {author} {\bibfnamefont
  {T.}~\bibnamefont {Walker}}, \ and\ \bibinfo {author} {\bibfnamefont
  {M.}~\bibnamefont {Saffman}},\ }\href@noop {} {\bibfield  {journal} {\bibinfo
   {journal} {Physical Review Letters}\ }\textbf {\bibinfo {volume} {104}},\
  \bibinfo {pages} {010503} (\bibinfo {year} {2010})}\BibitemShut {NoStop}%
\bibitem [{\citenamefont {Buluta}\ and\ \citenamefont
  {Nori}(2009)}]{buluta2009}%
  \BibitemOpen
  \bibfield  {author} {\bibinfo {author} {\bibfnamefont {I.}~\bibnamefont
  {Buluta}}\ and\ \bibinfo {author} {\bibfnamefont {F.}~\bibnamefont {Nori}},\
  }\href@noop {} {\bibfield  {journal} {\bibinfo  {journal} {Science}\ }\textbf
  {\bibinfo {volume} {326}},\ \bibinfo {pages} {108} (\bibinfo {year}
  {2009})}\BibitemShut {NoStop}%
\bibitem [{\citenamefont {Weimer}\ \emph {et~al.}(2010)\citenamefont {Weimer},
  \citenamefont {M{\"u}ller}, \citenamefont {Lesanovsky}, \citenamefont
  {Zoller},\ and\ \citenamefont {B{\"u}chler}}]{weimer2010}%
  \BibitemOpen
  \bibfield  {author} {\bibinfo {author} {\bibfnamefont {H.}~\bibnamefont
  {Weimer}}, \bibinfo {author} {\bibfnamefont {M.}~\bibnamefont {M{\"u}ller}},
  \bibinfo {author} {\bibfnamefont {I.}~\bibnamefont {Lesanovsky}}, \bibinfo
  {author} {\bibfnamefont {P.}~\bibnamefont {Zoller}}, \ and\ \bibinfo {author}
  {\bibfnamefont {H.~P.}\ \bibnamefont {B{\"u}chler}},\ }\href@noop {}
  {\bibfield  {journal} {\bibinfo  {journal} {Nature Physics}\ }\textbf
  {\bibinfo {volume} {6}},\ \bibinfo {pages} {382} (\bibinfo {year}
  {2010})}\BibitemShut {NoStop}%
\bibitem [{\citenamefont {Lanyon}\ \emph {et~al.}(2011)\citenamefont {Lanyon},
  \citenamefont {Hempel}, \citenamefont {Nigg}, \citenamefont {M{\"u}ller},
  \citenamefont {Gerritsma}, \citenamefont {Z{\"a}hringer}, \citenamefont
  {Schindler}, \citenamefont {Barreiro}, \citenamefont {Rambach}, \citenamefont
  {Kirchmair} \emph {et~al.}}]{lanyon2011}%
  \BibitemOpen
  \bibfield  {author} {\bibinfo {author} {\bibfnamefont {B.~P.}\ \bibnamefont
  {Lanyon}}, \bibinfo {author} {\bibfnamefont {C.}~\bibnamefont {Hempel}},
  \bibinfo {author} {\bibfnamefont {D.}~\bibnamefont {Nigg}}, \bibinfo {author}
  {\bibfnamefont {M.}~\bibnamefont {M{\"u}ller}}, \bibinfo {author}
  {\bibfnamefont {R.}~\bibnamefont {Gerritsma}}, \bibinfo {author}
  {\bibfnamefont {F.}~\bibnamefont {Z{\"a}hringer}}, \bibinfo {author}
  {\bibfnamefont {P.}~\bibnamefont {Schindler}}, \bibinfo {author}
  {\bibfnamefont {J.}~\bibnamefont {Barreiro}}, \bibinfo {author}
  {\bibfnamefont {M.}~\bibnamefont {Rambach}}, \bibinfo {author} {\bibfnamefont
  {G.}~\bibnamefont {Kirchmair}},  \emph {et~al.},\ }\href@noop {} {\bibfield
  {journal} {\bibinfo  {journal} {Science}\ }\textbf {\bibinfo {volume}
  {334}},\ \bibinfo {pages} {57} (\bibinfo {year} {2011})}\BibitemShut
  {NoStop}%
\bibitem [{\citenamefont {Salath\'e}\ \emph {et~al.}(2015)\citenamefont
  {Salath\'e}, \citenamefont {Mondal}, \citenamefont {Oppliger}, \citenamefont
  {Heinsoo}, \citenamefont {Kurpiers}, \citenamefont
  {Poto\ifmmode~\check{c}\else \v{c}\fi{}nik}, \citenamefont {Mezzacapo},
  \citenamefont {Las~Heras}, \citenamefont {Lamata}, \citenamefont {Solano},
  \citenamefont {Filipp},\ and\ \citenamefont {Wallraff}}]{Salathe2015}%
  \BibitemOpen
  \bibfield  {author} {\bibinfo {author} {\bibfnamefont {Y.}~\bibnamefont
  {Salath\'e}}, \bibinfo {author} {\bibfnamefont {M.}~\bibnamefont {Mondal}},
  \bibinfo {author} {\bibfnamefont {M.}~\bibnamefont {Oppliger}}, \bibinfo
  {author} {\bibfnamefont {J.}~\bibnamefont {Heinsoo}}, \bibinfo {author}
  {\bibfnamefont {P.}~\bibnamefont {Kurpiers}}, \bibinfo {author}
  {\bibfnamefont {A.}~\bibnamefont {Poto\ifmmode~\check{c}\else
  \v{c}\fi{}nik}}, \bibinfo {author} {\bibfnamefont {A.}~\bibnamefont
  {Mezzacapo}}, \bibinfo {author} {\bibfnamefont {U.}~\bibnamefont
  {Las~Heras}}, \bibinfo {author} {\bibfnamefont {L.}~\bibnamefont {Lamata}},
  \bibinfo {author} {\bibfnamefont {E.}~\bibnamefont {Solano}}, \bibinfo
  {author} {\bibfnamefont {S.}~\bibnamefont {Filipp}}, \ and\ \bibinfo {author}
  {\bibfnamefont {A.}~\bibnamefont {Wallraff}},\ }\href {\doibase
  10.1103/PhysRevX.5.021027} {\bibfield  {journal} {\bibinfo  {journal} {Phys.
  Rev. X}\ }\textbf {\bibinfo {volume} {5}},\ \bibinfo {pages} {021027}
  (\bibinfo {year} {2015})}\BibitemShut {NoStop}%
\bibitem [{\citenamefont {Barends}\ \emph {et~al.}(2015)\citenamefont
  {Barends}, \citenamefont {Lamata}, \citenamefont {Kelly}, \citenamefont
  {Garc{\'\i}a-{\'A}lvarez}, \citenamefont {Fowler}, \citenamefont {Megrant},
  \citenamefont {Jeffrey}, \citenamefont {White}, \citenamefont {Sank},
  \citenamefont {Mutus} \emph {et~al.}}]{barends2015}%
  \BibitemOpen
  \bibfield  {author} {\bibinfo {author} {\bibfnamefont {R.}~\bibnamefont
  {Barends}}, \bibinfo {author} {\bibfnamefont {L.}~\bibnamefont {Lamata}},
  \bibinfo {author} {\bibfnamefont {J.}~\bibnamefont {Kelly}}, \bibinfo
  {author} {\bibfnamefont {L.}~\bibnamefont {Garc{\'\i}a-{\'A}lvarez}},
  \bibinfo {author} {\bibfnamefont {A.}~\bibnamefont {Fowler}}, \bibinfo
  {author} {\bibfnamefont {A.}~\bibnamefont {Megrant}}, \bibinfo {author}
  {\bibfnamefont {E.}~\bibnamefont {Jeffrey}}, \bibinfo {author} {\bibfnamefont
  {T.}~\bibnamefont {White}}, \bibinfo {author} {\bibfnamefont
  {D.}~\bibnamefont {Sank}}, \bibinfo {author} {\bibfnamefont {J.}~\bibnamefont
  {Mutus}},  \emph {et~al.},\ }\href@noop {} {\bibfield  {journal} {\bibinfo
  {journal} {Nature communications}\ }\textbf {\bibinfo {volume} {6}},\
  \bibinfo {pages} {7654} (\bibinfo {year} {2015})}\BibitemShut {NoStop}%
\bibitem [{\citenamefont {Gr{\"o}ssing}\ and\ \citenamefont
  {Zeilinger}(1988)}]{grossing1988}%
  \BibitemOpen
  \bibfield  {author} {\bibinfo {author} {\bibfnamefont {G.}~\bibnamefont
  {Gr{\"o}ssing}}\ and\ \bibinfo {author} {\bibfnamefont {A.}~\bibnamefont
  {Zeilinger}},\ }\href@noop {} {\bibfield  {journal} {\bibinfo  {journal}
  {Complex systems}\ }\textbf {\bibinfo {volume} {2}},\ \bibinfo {pages} {197}
  (\bibinfo {year} {1988})}\BibitemShut {NoStop}%
\bibitem [{\citenamefont {Lent}\ \emph {et~al.}(1993)\citenamefont {Lent},
  \citenamefont {Tougaw}, \citenamefont {Porod},\ and\ \citenamefont
  {Bernstein}}]{lent1993}%
  \BibitemOpen
  \bibfield  {author} {\bibinfo {author} {\bibfnamefont {C.~S.}\ \bibnamefont
  {Lent}}, \bibinfo {author} {\bibfnamefont {P.~D.}\ \bibnamefont {Tougaw}},
  \bibinfo {author} {\bibfnamefont {W.}~\bibnamefont {Porod}}, \ and\ \bibinfo
  {author} {\bibfnamefont {G.~H.}\ \bibnamefont {Bernstein}},\ }\href@noop {}
  {\bibfield  {journal} {\bibinfo  {journal} {Nanotechnology}\ }\textbf
  {\bibinfo {volume} {4}},\ \bibinfo {pages} {49} (\bibinfo {year}
  {1993})}\BibitemShut {NoStop}%
\bibitem [{\citenamefont {Meyer}(1996)}]{meyer1996}%
  \BibitemOpen
  \bibfield  {author} {\bibinfo {author} {\bibfnamefont {D.~A.}\ \bibnamefont
  {Meyer}},\ }\href@noop {} {\bibfield  {journal} {\bibinfo  {journal} {Journal
  of Statistical Physics}\ }\textbf {\bibinfo {volume} {85}},\ \bibinfo {pages}
  {551} (\bibinfo {year} {1996})}\BibitemShut {NoStop}%
\bibitem [{\citenamefont {G{\"u}tschow}\ \emph {et~al.}(2010)\citenamefont
  {G{\"u}tschow}, \citenamefont {Uphoff}, \citenamefont {Werner},\ and\
  \citenamefont {Zimbor{\'a}s}}]{gutschow2010}%
  \BibitemOpen
  \bibfield  {author} {\bibinfo {author} {\bibfnamefont {J.}~\bibnamefont
  {G{\"u}tschow}}, \bibinfo {author} {\bibfnamefont {S.}~\bibnamefont
  {Uphoff}}, \bibinfo {author} {\bibfnamefont {R.~F.}\ \bibnamefont {Werner}},
  \ and\ \bibinfo {author} {\bibfnamefont {Z.}~\bibnamefont {Zimbor{\'a}s}},\
  }\href@noop {} {\bibfield  {journal} {\bibinfo  {journal} {Journal of
  Mathematical Physics}\ }\textbf {\bibinfo {volume} {51}},\ \bibinfo {pages}
  {015203} (\bibinfo {year} {2010})}\BibitemShut {NoStop}%
\bibitem [{\citenamefont {Alonso-Sanz}(2014)}]{alonso2014}%
  \BibitemOpen
  \bibfield  {author} {\bibinfo {author} {\bibfnamefont {R.}~\bibnamefont
  {Alonso-Sanz}},\ }\href@noop {} {\bibfield  {journal} {\bibinfo  {journal}
  {Proc. R. Soc. A}\ }\textbf {\bibinfo {volume} {470}},\ \bibinfo {pages}
  {20130793} (\bibinfo {year} {2014})}\BibitemShut {NoStop}%
\bibitem [{\citenamefont {Prosen}\ and\ \citenamefont
  {Mej{\'\i}a-Monasterio}(2016)}]{prosen2016}%
  \BibitemOpen
  \bibfield  {author} {\bibinfo {author} {\bibfnamefont {T.}~\bibnamefont
  {Prosen}}\ and\ \bibinfo {author} {\bibfnamefont {C.}~\bibnamefont
  {Mej{\'\i}a-Monasterio}},\ }\href@noop {} {\bibfield  {journal} {\bibinfo
  {journal} {Journal of Physics A}\ }\textbf {\bibinfo {volume} {49}},\
  \bibinfo {pages} {185003} (\bibinfo {year} {2016})}\BibitemShut {NoStop}%
\bibitem [{\citenamefont {Gopalakrishnan}\ and\ \citenamefont
  {Zakirov}(2018)}]{gopalakrishnan2018}%
  \BibitemOpen
  \bibfield  {author} {\bibinfo {author} {\bibfnamefont {S.}~\bibnamefont
  {Gopalakrishnan}}\ and\ \bibinfo {author} {\bibfnamefont {B.}~\bibnamefont
  {Zakirov}},\ }\href@noop {} {\bibfield  {journal} {\bibinfo  {journal}
  {arXiv:1802.07729}\ } (\bibinfo {year} {2018})}\BibitemShut {NoStop}%
\bibitem [{\citenamefont {Lewenstein}(1994)}]{lewenstein1994}%
  \BibitemOpen
  \bibfield  {author} {\bibinfo {author} {\bibfnamefont {M.}~\bibnamefont
  {Lewenstein}},\ }\href@noop {} {\bibfield  {journal} {\bibinfo  {journal}
  {Journal of Modern Optics}\ }\textbf {\bibinfo {volume} {41}},\ \bibinfo
  {pages} {2491} (\bibinfo {year} {1994})}\BibitemShut {NoStop}%
\bibitem [{\citenamefont {Torrontegui}\ and\ \citenamefont
  {Garcia-Ripoll}(2018)}]{torrontegui2018}%
  \BibitemOpen
  \bibfield  {author} {\bibinfo {author} {\bibfnamefont {E.}~\bibnamefont
  {Torrontegui}}\ and\ \bibinfo {author} {\bibfnamefont {J.}~\bibnamefont
  {Garcia-Ripoll}},\ }\href@noop {} {\bibfield  {journal} {\bibinfo  {journal}
  {arXiv:1801.00934}\ } (\bibinfo {year} {2018})}\BibitemShut {NoStop}%
\bibitem [{\citenamefont {Ritort}\ and\ \citenamefont
  {Sollich}(2003)}]{Ritort2003}%
  \BibitemOpen
  \bibfield  {author} {\bibinfo {author} {\bibfnamefont {F.}~\bibnamefont
  {Ritort}}\ and\ \bibinfo {author} {\bibfnamefont {P.}~\bibnamefont
  {Sollich}},\ }\href {\doibase 10.1080/0001873031000093582} {\bibfield
  {journal} {\bibinfo  {journal} {Adv. Phys.}\ }\textbf {\bibinfo {volume}
  {52}},\ \bibinfo {pages} {219} (\bibinfo {year} {2003})}\BibitemShut
  {NoStop}%
\bibitem [{\citenamefont {Garrahan}\ and\ \citenamefont
  {Chandler}(2002)}]{Garrahan2002}%
  \BibitemOpen
  \bibfield  {author} {\bibinfo {author} {\bibfnamefont {J.~P.}\ \bibnamefont
  {Garrahan}}\ and\ \bibinfo {author} {\bibfnamefont {D.}~\bibnamefont
  {Chandler}},\ }\href@noop {} {\bibfield  {journal} {\bibinfo  {journal}
  {Phys. Rev. Lett.}\ }\textbf {\bibinfo {volume} {89}},\ \bibinfo {pages}
  {035704} (\bibinfo {year} {2002})}\BibitemShut {NoStop}%
\bibitem [{\citenamefont {Olmos}\ \emph {et~al.}(2012)\citenamefont {Olmos},
  \citenamefont {Lesanovsky},\ and\ \citenamefont {Garrahan}}]{Olmos2012}%
  \BibitemOpen
  \bibfield  {author} {\bibinfo {author} {\bibfnamefont {B.}~\bibnamefont
  {Olmos}}, \bibinfo {author} {\bibfnamefont {I.}~\bibnamefont {Lesanovsky}}, \
  and\ \bibinfo {author} {\bibfnamefont {J.~P.}\ \bibnamefont {Garrahan}},\
  }\href {\doibase 10.1103/PhysRevLett.109.020403} {\bibfield  {journal}
  {\bibinfo  {journal} {Phys. Rev. Lett.}\ }\textbf {\bibinfo {volume} {109}},\
  \bibinfo {pages} {020403} (\bibinfo {year} {2012})}\BibitemShut {NoStop}%
\bibitem [{\citenamefont {van Horssen}\ \emph {et~al.}(2015)\citenamefont {van
  Horssen}, \citenamefont {Levi},\ and\ \citenamefont
  {Garrahan}}]{Horssen2015}%
  \BibitemOpen
  \bibfield  {author} {\bibinfo {author} {\bibfnamefont {M.}~\bibnamefont {van
  Horssen}}, \bibinfo {author} {\bibfnamefont {E.}~\bibnamefont {Levi}}, \ and\
  \bibinfo {author} {\bibfnamefont {J.~P.}\ \bibnamefont {Garrahan}},\ }\href
  {\doibase 10.1103/PhysRevB.92.100305} {\bibfield  {journal} {\bibinfo
  {journal} {Phys. Rev. B}\ }\textbf {\bibinfo {volume} {92}},\ \bibinfo
  {pages} {100305} (\bibinfo {year} {2015})}\BibitemShut {NoStop}%
\bibitem [{\citenamefont {Lan}\ \emph {et~al.}(2017)\citenamefont {Lan},
  \citenamefont {van Horssen}, \citenamefont {Powell},\ and\ \citenamefont
  {Garrahan}}]{Lan2017}%
  \BibitemOpen
  \bibfield  {author} {\bibinfo {author} {\bibfnamefont {Z.}~\bibnamefont
  {Lan}}, \bibinfo {author} {\bibfnamefont {M.}~\bibnamefont {van Horssen}},
  \bibinfo {author} {\bibfnamefont {S.}~\bibnamefont {Powell}}, \ and\ \bibinfo
  {author} {\bibfnamefont {J.}~\bibnamefont {Garrahan}},\ }\href@noop {}
  {\bibfield  {journal} {\bibinfo  {journal} {arXiv:1706.02603}\ } (\bibinfo
  {year} {2017})}\BibitemShut {NoStop}%
\bibitem [{\citenamefont {Shiraishi}\ and\ \citenamefont
  {Mori}(2017)}]{Shiraishi2017}%
  \BibitemOpen
  \bibfield  {author} {\bibinfo {author} {\bibfnamefont {N.}~\bibnamefont
  {Shiraishi}}\ and\ \bibinfo {author} {\bibfnamefont {T.}~\bibnamefont
  {Mori}},\ }\href@noop {} {\bibfield  {journal} {\bibinfo  {journal}
  {arXiv:1702.08227}\ } (\bibinfo {year} {2017})}\BibitemShut {NoStop}%
\bibitem [{\citenamefont {Turner}\ \emph {et~al.}(2017)\citenamefont {Turner},
  \citenamefont {Michailidis}, \citenamefont {Abanin}, \citenamefont {Serbyn},\
  and\ \citenamefont {Papic}}]{Turner2017}%
  \BibitemOpen
  \bibfield  {author} {\bibinfo {author} {\bibfnamefont {C.~J.}\ \bibnamefont
  {Turner}}, \bibinfo {author} {\bibfnamefont {A.~A.}\ \bibnamefont
  {Michailidis}}, \bibinfo {author} {\bibfnamefont {D.~A.}\ \bibnamefont
  {Abanin}}, \bibinfo {author} {\bibfnamefont {M.}~\bibnamefont {Serbyn}}, \
  and\ \bibinfo {author} {\bibfnamefont {Z.}~\bibnamefont {Papic}},\
  }\href@noop {} {\bibfield  {journal} {\bibinfo  {journal} {arXiv:1711.03528}\
  } (\bibinfo {year} {2017})}\BibitemShut {NoStop}%
\bibitem [{\citenamefont {Nandkishore}\ and\ \citenamefont
  {Hermele}(2018)}]{Nandkishore2018}%
  \BibitemOpen
  \bibfield  {author} {\bibinfo {author} {\bibfnamefont {R.~M.}\ \bibnamefont
  {Nandkishore}}\ and\ \bibinfo {author} {\bibfnamefont {M.}~\bibnamefont
  {Hermele}},\ }\href@noop {} {\bibfield  {journal} {\bibinfo  {journal}
  {arXiv:1803.1196}\ } (\bibinfo {year} {2018})}\BibitemShut {NoStop}%
\bibitem [{\citenamefont {Whitelam}\ \emph {et~al.}(2004)\citenamefont
  {Whitelam}, \citenamefont {Berthier},\ and\ \citenamefont
  {Garrahan}}]{Whitelam2004}%
  \BibitemOpen
  \bibfield  {author} {\bibinfo {author} {\bibfnamefont {S.}~\bibnamefont
  {Whitelam}}, \bibinfo {author} {\bibfnamefont {L.}~\bibnamefont {Berthier}},
  \ and\ \bibinfo {author} {\bibfnamefont {J.}~\bibnamefont {Garrahan}},\
  }\href@noop {} {\bibfield  {journal} {\bibinfo  {journal} {Phys. Rev. Lett.}\
  }\textbf {\bibinfo {volume} {92}},\ \bibinfo {pages} {185705} (\bibinfo
  {year} {2004})}\BibitemShut {NoStop}%
\bibitem [{\citenamefont {Jack}\ \emph {et~al.}(2006)\citenamefont {Jack},
  \citenamefont {Mayer},\ and\ \citenamefont {Sollich}}]{Jack2006}%
  \BibitemOpen
  \bibfield  {author} {\bibinfo {author} {\bibfnamefont {R.~L.}\ \bibnamefont
  {Jack}}, \bibinfo {author} {\bibfnamefont {P.}~\bibnamefont {Mayer}}, \ and\
  \bibinfo {author} {\bibfnamefont {P.}~\bibnamefont {Sollich}},\ }\href@noop
  {} {\bibfield  {journal} {\bibinfo  {journal} {J. Stat. Mech.}\ ,\ \bibinfo
  {pages} {P03006}} (\bibinfo {year} {2006})}\BibitemShut {NoStop}%
\bibitem [{\citenamefont {Hinrichsen}(2000)}]{hinrichsen2000}%
  \BibitemOpen
  \bibfield  {author} {\bibinfo {author} {\bibfnamefont {H.}~\bibnamefont
  {Hinrichsen}},\ }\href@noop {} {\bibfield  {journal} {\bibinfo  {journal}
  {Adv. Phys.}\ }\textbf {\bibinfo {volume} {49}},\ \bibinfo {pages} {815}
  (\bibinfo {year} {2000})}\BibitemShut {NoStop}%
\bibitem [{\citenamefont {Domany}\ and\ \citenamefont
  {Kinzel}(1984)}]{domany1984}%
  \BibitemOpen
  \bibfield  {author} {\bibinfo {author} {\bibfnamefont {E.}~\bibnamefont
  {Domany}}\ and\ \bibinfo {author} {\bibfnamefont {W.}~\bibnamefont
  {Kinzel}},\ }\href@noop {} {\bibfield  {journal} {\bibinfo  {journal}
  {Physical review letters}\ }\textbf {\bibinfo {volume} {53}},\ \bibinfo
  {pages} {311} (\bibinfo {year} {1984})}\BibitemShut {NoStop}%
\bibitem [{\citenamefont {Girolami}\ \emph {et~al.}(2013)\citenamefont
  {Girolami}, \citenamefont {Tufarelli},\ and\ \citenamefont
  {Adesso}}]{girolami2013}%
  \BibitemOpen
  \bibfield  {author} {\bibinfo {author} {\bibfnamefont {D.}~\bibnamefont
  {Girolami}}, \bibinfo {author} {\bibfnamefont {T.}~\bibnamefont {Tufarelli}},
  \ and\ \bibinfo {author} {\bibfnamefont {G.}~\bibnamefont {Adesso}},\
  }\href@noop {} {\bibfield  {journal} {\bibinfo  {journal} {Physical Review
  Letters}\ }\textbf {\bibinfo {volume} {110}},\ \bibinfo {pages} {240402}
  (\bibinfo {year} {2013})}\BibitemShut {NoStop}%
\bibitem [{\citenamefont {Zurek}(2000)}]{Zurek2000}%
  \BibitemOpen
  \bibfield  {author} {\bibinfo {author} {\bibfnamefont {W.}~\bibnamefont
  {Zurek}},\ }\href {\doibase
  10.1002/1521-3889(200011)9:11/12<855::AID-ANDP855>3.0.CO;2-K} {\bibfield
  {journal} {\bibinfo  {journal} {Annalen der Physik}\ }\textbf {\bibinfo
  {volume} {9}},\ \bibinfo {pages} {855} (\bibinfo {year} {2000})}\BibitemShut
  {NoStop}%
\bibitem [{\citenamefont {Ollivier}\ and\ \citenamefont
  {Zurek}(2001)}]{Ollivier2001}%
  \BibitemOpen
  \bibfield  {author} {\bibinfo {author} {\bibfnamefont {H.}~\bibnamefont
  {Ollivier}}\ and\ \bibinfo {author} {\bibfnamefont {W.~H.}\ \bibnamefont
  {Zurek}},\ }\href {\doibase 10.1103/PhysRevLett.88.017901} {\bibfield
  {journal} {\bibinfo  {journal} {Phys. Rev. Lett.}\ }\textbf {\bibinfo
  {volume} {88}},\ \bibinfo {pages} {017901} (\bibinfo {year}
  {2001})}\BibitemShut {NoStop}%
\bibitem [{\citenamefont {Henderson}\ and\ \citenamefont
  {Vedral}(2001)}]{Henderson2001}%
  \BibitemOpen
  \bibfield  {author} {\bibinfo {author} {\bibfnamefont {L.}~\bibnamefont
  {Henderson}}\ and\ \bibinfo {author} {\bibfnamefont {V.}~\bibnamefont
  {Vedral}},\ }\href {http://stacks.iop.org/0305-4470/34/i=35/a=315} {\bibfield
   {journal} {\bibinfo  {journal} {Journal of Physics A}\ }\textbf {\bibinfo
  {volume} {34}},\ \bibinfo {pages} {6899} (\bibinfo {year}
  {2001})}\BibitemShut {NoStop}%
\bibitem [{\citenamefont {Ostmann}\ \emph {et~al.}(2017)\citenamefont
  {Ostmann}, \citenamefont {Min{\'a}{\v{r}}}, \citenamefont {Marcuzzi},
  \citenamefont {Levi},\ and\ \citenamefont {Lesanovsky}}]{ostmann2017}%
  \BibitemOpen
  \bibfield  {author} {\bibinfo {author} {\bibfnamefont {M.}~\bibnamefont
  {Ostmann}}, \bibinfo {author} {\bibfnamefont {J.}~\bibnamefont
  {Min{\'a}{\v{r}}}}, \bibinfo {author} {\bibfnamefont {M.}~\bibnamefont
  {Marcuzzi}}, \bibinfo {author} {\bibfnamefont {E.}~\bibnamefont {Levi}}, \
  and\ \bibinfo {author} {\bibfnamefont {I.}~\bibnamefont {Lesanovsky}},\
  }\href@noop {} {\bibfield  {journal} {\bibinfo  {journal} {New Journal of
  Physics}\ }\textbf {\bibinfo {volume} {19}},\ \bibinfo {pages} {123015}
  (\bibinfo {year} {2017})}\BibitemShut {NoStop}%
\bibitem [{\citenamefont {Essam}(1989)}]{essam1989}%
  \BibitemOpen
  \bibfield  {author} {\bibinfo {author} {\bibfnamefont {J.}~\bibnamefont
  {Essam}},\ }\href@noop {} {\bibfield  {journal} {\bibinfo  {journal} {Journal
  of Physics A}\ }\textbf {\bibinfo {volume} {22}},\ \bibinfo {pages} {4927}
  (\bibinfo {year} {1989})}\BibitemShut {NoStop}%
\bibitem [{\citenamefont {L{\"u}beck}(2006)}]{lubeck2006}%
  \BibitemOpen
  \bibfield  {author} {\bibinfo {author} {\bibfnamefont {S.}~\bibnamefont
  {L{\"u}beck}},\ }\href@noop {} {\bibfield  {journal} {\bibinfo  {journal}
  {Journal of Statistical Physics}\ }\textbf {\bibinfo {volume} {123}},\
  \bibinfo {pages} {193} (\bibinfo {year} {2006})}\BibitemShut {NoStop}%
\bibitem [{\citenamefont {Roscher}\ \emph {et~al.}(2018)\citenamefont
  {Roscher}, \citenamefont {Diehl},\ and\ \citenamefont
  {Buchhold}}]{roscher2018}%
  \BibitemOpen
  \bibfield  {author} {\bibinfo {author} {\bibfnamefont {D.}~\bibnamefont
  {Roscher}}, \bibinfo {author} {\bibfnamefont {S.}~\bibnamefont {Diehl}}, \
  and\ \bibinfo {author} {\bibfnamefont {M.}~\bibnamefont {Buchhold}},\
  }\href@noop {} {\bibfield  {journal} {\bibinfo  {journal} {arXiv:1803.08514}\
  } (\bibinfo {year} {2018})}\BibitemShut {NoStop}%
\end{thebibliography}

%

\end{document}